\documentstyle[11pt,paspconf,epsfig]{article}
 
\def\>{$>$}
\def\<{$<$}

\def\simlt{\lower.5ex\hbox{$\; \buildrel < \over \sim \;$}}
\def\simgt{\lower.5ex\hbox{$\; \buildrel > \over \sim \;$}}
\def\ch2{$\chi^{2}$}

\begin{document}

\def\dm{\dot{m}}
\def\dM{\dot{M}}
\def\LE{L_{\rm E}}
\def\dmcr{\dot{m}_{\rm cr}}
\def\te{\theta_e}
\def\tp{\theta_p}
\def\ep{\varepsilon}
\def\mus{\mu_{\rm s}}
\def\sT{\sigma_{\rm T}}
\def\OmK{\Omega_{\rm K}}
\def\vK{v_{\rm K}}
\def\tT{\tau_{\rm T}}
\def\trf{t_{r\varphi}}
\def\Trf{T_{r\varphi}}
\def\beq{\begin{equation}}
\def\eeq{\end{equation}}
\def\rms{r_{\rm ms}}
\def\rin{r_{\rm in}}
\def\bin{b_{\rm in}}
\def\dd{{\rm d}}
\def\rtr{r_{\rm tr}}

\title{Accretion Disk Models}
 
\author{Andrei M. Beloborodov\altaffilmark{1}}
\affil{Stockholm Observatory, Saltsj\"obaden, S-133 36, Sweden}

\altaffiltext{1}{Also at Astro-Space Center of Lebedev Physical Institute,
Profsoyuznaya 84/32, 117810 Moscow, Russia}

\begin{abstract}
Models of black hole accretion disks are reviewed, 
with an emphasis on the theory of hard X-ray production.
The following models are considered: i) standard, ii) super-critical, 
iii) two-temperature, and iv) disk+corona.
New developments have recently been made in hydrodynamical models of accretion
%have recently been developed, 
and in phenomenological radiative models
fitting the observed X-ray spectra.
Attempts to unify the two approaches are only partly successful.
%and a physical picture of 
%hard X-ray production 
%energy release near the black hole 
%is still not established.
\end{abstract}


%%%%%%%%%%%%%%%%%%%%%%%%%%%%%%%%%%%%%%%%%%%%%%%%%%%%%%%%%%%%%%%%%%%%%%%%%%%%%%

\section{Introduction}

Accretion disks around black holes are efficient radiators which can convert
a sizable fraction of gas rest mass energy into radiation. The disk forms if
the specific angular momentum of the accreting gas $j\gg r_gc$, where 
$r_g=2GM/c^2$
is the gravitational radius of the black hole. Then the radial infall is stopped
by the centrifugal potential barrier at a radius $r_d\sim j^2/r_gc^2$, the gas
cools and forms a ring rotating with Keplerian velocity.
Further accretion is possible only if some mechanism redistributes 
angular momentum, allowing gas to spiral toward the black hole.
A plausible mechanism is provided by MHD turbulence that  
develops due to the differential character of Keplerian rotation
(see Balbus \& Hawley 1998 for a review). Turbulent pulsations force each gas 
element to diffuse from one Keplerian orbit to another, and a ring of gas 
spreads out to form an extended disk. At the inner edge of 
the disk gas is absorbed by the black hole. The diffusion problem with the 
absorbing inner boundary has stationary solutions, each specified by 
a net accretion rate, $\dM$. 
The redistribution of angular momentum is accompanied by a viscous heating. 
As a result, the binding energy of spiraling gas is dissipated into heat
and can be radiated.

In this review, we concentrate on steady disk models
and their applications to active galactic nuclei (AGNs) and galactic black
hole candidates (GBHs). 
Observed spectra of the black hole sources indicate that accretion disks 
usually do {\it not} radiate as a blackbody. In particular, a huge luminosity 
comes out in hard X-rays. 
The origin of the hard X-ray component has been in the focus of
interest since the component was detected, and it remains as a challenge 
for theoretical models of accretion disks. 
We start with the standard disk model (Section 2) and then discuss the concept 
of advective disks (Section 3). In Section 4, we discuss 
super-critical disk models.
In Section 5, two-temperature models are reviewed. Finally, in Section 6,
we summarize new developments concerning the disk-corona model.

%###########################################################################

\section{The standard model}

The standard model (see Pringle 1981 for a review) 
provides a commonly used simple description of an accretion disk. 
The model considers a thin gaseous disk of surface density $\Sigma$ and 
density $\rho=\Sigma/2H$, where $H$ is the disk height-scale.
The disk rotates around the black hole with a Keplerian angular velocity,
$\OmK=(GM/r^3)^{1/2}$. $H$ is regulated by the pressure in the disk, $p$,
which supports the gas against the vertical component of gravity.
The vertical balance reads
\beq
\frac{H}{r}=\frac{c_s}{\vK},
\eeq
where $c_s^2\equiv p/\rho$ and $\vK=\OmK r$. 
The matter in the disk gradually drifts inward due to a viscous stress
$\trf=\nu\rho r (\dd\OmK/\dd r)$ where $\nu$ is a kinematic viscosity coefficient.
The stress is a fraction of the pressure, $\trf=\alpha p$, 
where $\alpha<1$ is assumed to be constant (Shakura 1972).
Equivalently, $\nu=(2/3)\alpha c_s H$.
The inward drift is governed by the angular momentum equation, 
\beq
  vr\Sigma\frac{\dd}{\dd r}(\OmK r^2)=\frac{\dd}{\dd r}(\Trf r^2), \qquad
% \mbox{where}\qquad 
   \Trf\equiv 2H\trf=\frac{3}{2}\nu\Sigma\OmK.
\eeq
Here $v$ is the accretion velocity. In a steady disk,
\beq
   2\pi rv\Sigma=\dM=const,
\eeq
which allows the angular momentum equation to be integrated,
\beq
   \Trf=\frac{\dM\OmK}{2\pi}\left(1-\frac{C}{\sqrt{r}}\right).
\eeq
The constant $C$ is determined by an inner boundary condition. 
In the very vicinity of the black hole, Keplerian rotation becomes unstable 
and gas falls into the black hole with a constant angular momentum.
The radius of the marginally stable Keplerian orbit is $\rms=3r_g$ for a 
Schwarzschild black hole. 
The standard model treats the transition to free fall 
by placing the inner disk boundary at $\rin=\rms$ and by assuming 
$\trf=0$ at $r=\rin$. Then $C=\sqrt{3r_g}$.

The accretion velocity may be expressed from (3) and (4) as
\beq
   v=\frac{3}{2}\frac{\nu}{r}S^{-1}=\alpha\frac{c_s^2}{\vK}S^{-1},
\qquad S\equiv 1-\left(\frac{3r_g}{r}\right)^{1/2}.
\eeq

The viscous heating rate (per unit area of the disk) is
\beq
  F^+=\Trf r\frac{\dd\OmK}{\dd r}=\frac{3}{2}\Trf\OmK=\frac{3\dM}{4\pi}\OmK^2S.
\eeq
The standard model assumes that the bulk of the dissipated energy 
is radiated away locally, i.e., the local heating rate, $F^+$, is balanced by 
the radiative losses from the two faces of the disk, $2F=F^-$. 
The disk therefore remains cold and thin, $c_s\ll \vK$, $H\ll r$, and
accretion is slow, $v\ll \vK$. 
%The assumption $\Omega=\OmK$ is self-consistent with high 
%accuracy $\sim (c_s/\vK)^2$.

As the binding energy is radiated away, the radiative efficiency of accretion
is $\eta=\bin/c^2$, where $\bin$ is the specific binding energy at the inner edge.
In the simplest version (eqs. [1-6]), a Newtonian gravitational
field is assumed, with the potential $GM/r$. Then $\bin=GM/2\rin=c^2/12$.
A relativistic version (Novikov \& Thorne 1973; Page \& Thorne 1974) yields
$0.06<\eta<0.42$ depending on the black hole spin. 
The highest efficiency is achieved for a maximally rotating black hole.
Then the stable Keplerian disk spreads deep into the potential well, 
$\rms\rightarrow r_g/2$, and this results in a high binding energy at the 
inner edge, $\bin=c^2(1-1/\sqrt{3})\approx 0.42c^2$.

The set of disk structure equations (1), (3), (5), and (6) should be completed 
by an equation of state and an equation specifying a mechanism of radiative 
cooling.
In particular, if (i) the pressure in the disk is dominated by ionized gas
with a temperature $T$, so that $c_s^2\approx 2kT/m_p$, and (ii) the produced 
radiation is near thermodynamic equilibrium with the gas and gradually 
diffuses out of the disk, then $F\approx ca_rT^4/\sqrt{3}\tau=
(ca_rm_p^4/16\sqrt{3}k^4)c_s^8/\tau$, where $a_r=7.56\times 10^{-15}$ 
is the radiation constant and 
$2\tau=\sT\Sigma/m_p$ is the Thomson optical depth across the disk. 
It completes the set of equations and
the disk parameters can be expressed as analytical functions of radius, 
each model being specified by $\dM$ and $\alpha$ (Shakura \& Sunyaev 1973). 
We will hereafter express $\dM$ in a dimensionless form, 
$\dm=\dM c^2/\LE$, where $\LE=2\pi r_g m_pc^3/\sT$ is
the Eddington luminosity.

When $\dot{m}$ exceeds $\sim 0.1(\alpha M/M_\odot)^{-1/8}$
there appears a region in the disk where the radiation
pressure dominates the gas pressure. Then the disk thickness is determined by
the vertical radiation flux, $GMm_pH/r^3=F\sT/c$, and the set of 
equations may be closed by the condition $H=(3/4)\dm r_g S$.
The model then becomes viscously (Lightman \& Eardley 1974) and 
thermally (Shakura \& Sunyaev 1976) unstable. The instabilities are reviewed
by Julian Krolik in these proceedings.

Assuming that the disk radiates as a black body, $F=\sigma T_*^4$, one gets 
\beq
 T_*\approx 3.5\times 10^7 \dm^{1/4}\left(\frac{M}{M_\odot}\right)^{-1/4}
            \left(\frac{r}{r_g}\right)^{-3/4} S^{1/4}.
\eeq
A maximum surface temperature is $T_*\sim 10^5$ K for 
AGNs ($M\sim 10^7-10^9M_\odot$) and $T_*\sim 10^7$ K for 
GBHs ($M\sim 10 M_\odot$). It is the major success of the model that $T_*$
approximately corresponds to observed UV emission 
in AGNs and X-ray emission in GBHs. 
The hard X-ray component would, however, correspond to a temperature 
$\sim 10^9$ K. 
The standard model may yield such high temperatures only if the accretion rate 
is near the critical value, $\dmcr=\eta^{-1}$, that corresponds
to the critical Eddington luminosity.
Near-critical disks can strongly deviate from the blackbody state provided 
$\alpha$ is large (Shakura \& Sunyaev 1973). An overheating then 
occurs in the inner region of the disk where the inflow time-scale is shorter 
than the time-scale for relaxation to thermodynamic equilibrium.
This effect is discussed in more detail in Section 4.

%#############################################################################

\section{Advective disks: basic equations}

The key assumption of the standard model, that the dissipated energy is 
radiated away locally, is relaxed in the presently popular models of 
advection-dominated accretion flows (ADAFs, see Narayan, Mahadevan \& Quataert
1998 for a review). There are two types of ADAFs: super-critical (see Section 4) 
and two-temperature (see Section 5.2). 
In both models, a large fraction of the released energy is 
stored in the gas and advected into the black hole instead of being radiated 
away. 
%The advection-dominated regime has a low radiative efficiency and it  
%has two main applications: i) Low luminosity objects such as, e.g., Sgr A$^*$
%(Narayan et al. 1998b), and ii) Extremely luminous objects, with luminosities 
%near the Eddington limit and super-critical (super-Eddington) accretion rates.
%The latter case  may apply to quasars (e.g., Szuszkiewicz, Malkan \& 
%Abramowicz 1996) and outbursts in GBHs. 
%
Advective disks are geometrically thick, and a two-dimensional (2D) approach
would be 
more adequate to the problem. Most of the models are, however, based on the 
vertically integrated (1D) equations due to their simplicity. 
The $\alpha$-parametrization of viscosity is used in the same manner as in the 
standard model, sometimes with modifications suppressing $\alpha$ at the
sonic radius to avoid the causality problem (Narayan 1992). 
By solving the 1D equations, one gets
typical parameters of the accreting gas as a function of radius. 

Advection implies a non-local character of the flow, so that one deals with 
a set of non-linear differential equations with boundary conditions.
In the innermost region, the flow passes through a sonic radius, $r_s$, where
the radial velocity $v$ exceeds the sound speed.
A regularity condition must be fulfilled 
for any steady transonic flow at $r=r_s$ (cf. Landau \& Lifshitz 1987). 
This condition reads $N=D=0$ where 
$N$ and $D$ are the numerator and denominator in the explicit expression for 
the derivative $\dd v/\dd r$. 
The regularity condition gives two extra equations that determine $r_s$ and
impose a connection between $\dM$ and the flow parameters at an external 
boundary.

The transonic character of accretion is related to a special feature of 
a black hole gravitational field: the effective potential for radial motion
with a given angular momentum has a maximum (see Misner, Thorne \& Wheeler 
1973). After passing the potential barrier, rotation cannot stop the inflow. 
%toward the black hole. 
%The inner (transonic) edge of the disk, where the 
%transition to the free fall occurs, resembles the $L_1$ point in a binary 
%system. 
%The relativistic nature of the gravitational field is
%crucial for advective disks and the Newtonian approximation may not be used. 
%
%The relativistic effects in the vicinity of a non-rotating black hole 
The relativistic nature of a black hole gravitational field
can be approximately described by 
the pseudo-Newtonian potential $\varphi=GM/(r-r_g)$ (Paczy\'nski \& Wiita 1980)
which is a good approximation for a Schwarzschild black hole.
Similar potentials have been
proposed to emulate the gravitational field of a rotating black hole (Artemova,
Bj\"ornsson \& Novikov 1996). 
%The 1D equations of advective disks in the pseudo-Newtonian potential 
The pseudo-Newtonian models were 
studied by Paczy\'nski \& Bisnovatyi-Kogan (1981) and in many subsequent works. 
Recently, fully relativistic 1D equations have been derived 
(Lasota 1994; Abramowicz et al. 1996).
We here summarize the relativistic equations with some corrections 
(Beloborodov, Abramowicz \& Novikov 1997).

First, we summarize notation.
The Kerr geometry is described by the metric tensor $g_{ij}$ in the 
Boyer-Lindquist coordinates $x^i = (t, r, \theta, \varphi)$
(given, e.g., in Misner et al. 1973). 
%The disc is assumed to lie at the equatorial plane of the Kerr geometry. 
The four-velocity of the accreting gas is  
$u^i =(u^t, u^r, u^{\theta}, u^{\varphi})$ with $u^\theta=0$ assumed in the 
1D model. The angular velocity of the gas is $\Omega = u^\varphi/u^t$
and its Lorentz factor is $\gamma = u^t(-g^{tt})^{-1/2}$ 
measured in the frame of local observers having zero angular momentum.
The vertically integrated parameters of the disk are:
surface rest mass density $\Sigma$, surface energy density $U=\Sigma c^2+\Pi$ 
($\Pi$ being the internal energy), and the vertically integrated pressure $P$. 
The dimensionless specific enthalpy is $\mu=(U+P)/\Sigma c^2$. 
$F^+$ is the rate of viscous heating, and 
$F^-$ is the local radiation flux from the two faces of the disk.
%All the thermodynamic quantities and $F^\pm$ 
$\Sigma,U,\Pi,P$, and $F^\pm$
are measured in the local comoving frame. 

The main equations of a steady disk express the general conservation laws 
discussed by Novikov \& Thorne (1973) and Page \& Thorne (1974) in the context
of the standard model. For advective disks,
the assumption $\Omega=\OmK^+$ is replaced by the radial momentum equation,
and the assumption $F^+=F^-$ is replaced by the energy equation including the
advection term. 
%We include in the equations the inertial mass
%associated with internal energy accumulated in the flow 
%Beloborodov, Abramowicz \& Novikov 1997).

%\newpage

\begin{eqnarray}
\nonumber
\mbox{\it Mass conservation:} \mbox{\hspace*{1.2cm}}
2\pi r c u^r \Sigma =-\dot{M} \mbox{\hspace*{15cm}}.
\end{eqnarray}

\begin{eqnarray}
\nonumber
\mbox{\it Angular momentum:}\qquad
  \frac{\dd}{\dd r}\left[\mu\left(\frac{\dot{M}u_\varphi}{2\pi}
  +2\nu\Sigma r \sigma_{~\varphi}^r\right)\right]=\frac{F^-}{c^2}\;ru_\varphi,
\mbox{\hspace*{3cm}}
\end{eqnarray}
%$\nu$ is a viscosity coefficient, $\sigma_{~\varphi}^r=
%(1/2)g^{rr}g_{\varphi\varphi}\sqrt{-g^{tt}} \gamma^3(d\Omega/dr)$ 
%is the shear. 
\begin{eqnarray}
\mbox{\hspace*{1.4cm}where\hspace*{1.2cm}} \sigma_{~\varphi}^r= 
  \frac{1}{2}\;g^{rr}g_{\varphi\varphi}\sqrt{-g^{tt}}
  \gamma^3\frac{\dd\Omega}{\dd r} \mbox{\hspace*{1.0cm}is the shear.} \nonumber
\end{eqnarray}

\begin{eqnarray}
\nonumber
\mbox{\it Energy:} \mbox{\hspace*{3.0cm}}
   F^+-F^-=cu^r\left(\frac{\dd\Pi}{\dd r}
   -\frac{\Pi+P}{\Sigma}\frac{\dd\Sigma}{\dd r}\right),
\mbox{\hspace*{15cm}}
\end{eqnarray}
\begin{eqnarray}
\nonumber
%  F^+=2\nu\Sigma \mu\;\sigma^2 c^2, \qquad
%  \sigma^2=\frac{1}{2} g^{rr}g_{\varphi\varphi}\left(-g^{tt}\right)
%  \gamma^4\left(\frac{d\Omega}{dr}\right)^2.  
\mbox{\hspace*{0.15cm}where\hspace*{1.2cm}}
F^+=\nu\Sigma \mu c^2 g^{rr}g_{\varphi\varphi}\left(-g^{tt}\right)
  \gamma^4\left(\frac{\dd\Omega}{\dd r}\right)^2.  
\end{eqnarray}

\begin{eqnarray}
\nonumber
\mbox{\it Radial momentum:}\qquad\;\;
   \frac{1}{2}\;\frac{\dd}{\dd r}\left(u_ru^r\right)=
   -\frac{1}{2}\;\frac{\partial g_{\varphi\varphi}}{\partial r}
   g^{tt}\gamma^2
   \left(\Omega-\OmK^+\right)\left(\Omega-\OmK^-\right)
\mbox{\hspace*{15cm}} \\
  -\frac{1}{c^2\Sigma \mu}\frac{\dd P}{\dd r}-\frac{F^+u_r}{c^3\Sigma \mu},
\mbox{\hspace*{15cm}}
\nonumber
\end{eqnarray}
where $\OmK^\pm$ are Keplerian angular velocities for the co-rotating
($+$) and counter-rotating ($-$) orbits
\begin{eqnarray}
\nonumber
\mbox{\hspace*{0.8cm}}\OmK^\pm=\pm\frac{c}{r(2r/r_g)^{1/2}\pm a_*r_g/2}.
\end{eqnarray}
Here $a_*\leq 1$ is the spin parameter of the black hole. 
%A maximally rotating hole has $a_*=1$.

The set of disk structure equations  gets closed when 
the viscosity $\nu$, the equation of state $P(\Pi)$, and the 
radiative cooling $F^-$ are specified. A standard 
prescription for viscosity is $\nu=\alpha c_sH$, where $\alpha$ is
a constant and $c_s=c(P/U)^{1/2}$ is the isothermal sound speed. 
The half-thickness of the disk, $H$, should be
estimated from the vertical balance condition. Near the black hole,
the tidal force compressing the disk in vertical direction depends on 
$\Omega$ (e.g., Abramowicz, Lanza \& Percival 1997). 
% Outside the sonic radius, the correction connected with deviations from
% Keplerian rotation does not exceed a few percent.
For $\Omega\approx\Omega^+_K$, which is a good approximation for 
the vertical balance, one gets (e.g., Riffert \& Herold 1995)
\begin{eqnarray}
\nonumber
   H^2=\frac{P}{U}\;\frac{2r^3}{r_gJ}, \qquad \mbox{where} \qquad
  J(a_*,r)=\frac{2(r^2-a_*r_g\sqrt{2r_gr}+0.75a_*^2r_g^2)}
  {2r^2-3r_gr+a_*r_g\sqrt{2r_gr}}
\end{eqnarray}
is a relativistic correction factor becoming unity at $r\gg r_g$.

\medskip
%\noindent
%{\it Global energy conservation}

The disk luminosity is related to $\dM$ by (Beloborodov et al. 1997)
\begin{eqnarray}
\nonumber
    L^-=-\frac{2\pi}{c}\int\limits_{r_s}^\infty u_t F^- r\dd r\approx\dot{M}c^2
    \left(1+\mu_{\rm in}\frac{u_t^{\rm in}}{c}\right),
\end{eqnarray}
where the index "in" refers to the inner transonic edge of the disk,
in particular, $\mu_{\rm in}$ is the dimensionless specific enpthalpy 
at the inner edge. The radiative efficiency of the disk equals
\begin{equation}
  \eta =\frac{L^-}{\dM c^2}= 1 - \mu_{\rm in}\left(1-\frac{\bin}{c^2}\right).
\end{equation}
Here, $\bin=c^2+u_t^{\rm in}c$ is the specific binding energy at the inner 
edge. In the standard model, $\mu_{\rm in}=1$ and $\eta=\bin/c^2$.
In the advective limit $\eta\rightarrow 0$, and hence
$\mu_{\rm in}\rightarrow (1-\bin/c^2)^{-1}$.
Note that one should not assume $\bin=0$ for advective disks.
In fact, the position and binding energy of the inner edge can be found
only by integrating the disk structure equations.
The difference $\mu_{\rm in}-1$ (which describes the relativistic increase of 
gas inertia due to stored heat) adjusts to keep $\eta\approx 0$.

%The advection effect results in that the luminosity is less than the total
%power released in the disc,
%\begin{eqnarray}
%\nonumber
%  L^+=-\frac{2\pi}{c}\int\limits_{r_s}^\infty u_t F^+ rdr.
%\end{eqnarray}
%The power "swallowed" by the black hole equals $L_{adv}=L^+-L^-$.

%The set of disk structure equations can be reduced to three ordinary
%differential equations to be solved numerically.

%###########################################################################

%%%%%%%%%%%%%%%%%%%%%%%%%%%%%%%%%%%%%%%%%%%%%%%%%%%%%%%%%%%%%%%%%%%%%%%%%

\section{Super-critical disks}

Large accretion rates are expected in the brightest objects, such as 
quasars or transient GBHs during outbursts. When the accretion rate approaches 
the critical value $\dmcr$ corresponding to the Eddington luminosity, 
the standard model becomes inconsistent. 
Inside some radius $r_t$, the produced radiation is trapped by the flow and 
advected into the black hole, as the inflow time-scale here is less 
than the time-scale of photon escape from the disk, $\sim \tau H/c$,
(Begelman \& Meier 1982). 
The trapping radius can be estimated from the standard model of Shakura \&
Sunyaev (1973) by 
comparing the radial flux of internal energy $3c_s^2\dM$ with the total
flux of radiation emitted outside $r$. 
One then gets $r_t\approx 7 r_g$ for $\dm=\dmcr=12$. 
The relative height of the standard disk, $H/r$, equals $\dm/27$ at the maximum.
Advection thus starts to become important before the accretion flow becomes 
quasi-spherical, and it may be approximately treated 
retaining the vertically-integrated approximation. 

Advection reduces the vertical radiation flux $F^-$ as compared to the 
standard model and, as a result, the disk thickness stays moderate at 
super-critical accretion rates (Abramowicz et al. 1988; Chen \& Taam 1993).
This made possible an extension of the 1D model to the super-Eddington
advection dominated regime, called ``slim'' accretion disk. 
The slim model may apply at
moderately super-critical accretion rates. In the limit $\dm\gg\dmcr$
there appears an extended region where the accretion flow has a positive 
Bernoulli constant, and 
the bulk of supplied gas may be pushed away by the radiation pressure to form 
a wind (Shakura \& Sunyaev 1973).
%; Blandford \& Begelman 1998). 
The behavior of the flow at $\dm\gg\dmcr$ still remains an open issue.
Detailed 2D hydrodynamical simulations might clarify whether gas mainly falls
into the black hole or flows out (Eggum, Coroniti \& Katz 1988; 
Igumenshchev \& Abramowicz 1998).
 
The relativistic slim disk has been calculated in Beloborodov (1998a) for both
non-rotating ($a_*=0$) and rapidly rotating ($a_*=0.998$) black holes. 
Like its pseudo-Newtonian counterpart,
the relativistic slim disk has a moderate height up to $\dm\sim 10\dmcr$
due to advection of the trapped radiation. 
%Here, $\dm_{cr}$ would correspond to $L=L_E$ in the relativistic standard 
%model: 
%$\dm_{cr}\approx 17.5$ for $a_*=0$ and $\dm_{cr}\approx 3.11$ for $a_*=0.998$. 
%
An important issue is the temperature of the accreting gas as the temperature
determines the emission spectrum. The simplest way to evaluate the temperature 
is by assuming that the gas is in thermodynamic equilibrium with the radiation
density in the disk, $w$. Then $T$ equals $T_{\rm eff}=(w/a_r)^{1/4}$ where,
$a_r$ is the radiation constant. This was
usually adopted in models of super-critical disks. The blackbody 
approximation, however, fails when $\alpha>0.03$ (Beloborodov 1998a). 
The gas then accretes so fast and has so low density that it is unable to 
reprocess the 
released energy into Planckian radiation. As a result, gas overheats
so that $T\gg T_{\rm eff}$. 

%%%%%%%%%%%%%%%%%%%%%%%%%%%%%%%%%%%%%%%%%%%%%%%%%%%%%%%%%%%%%%%%%%%%%%%%%%%
%Figure~\ref{fig-1}.
\begin{figure}
\centerline{
\psfig{figure=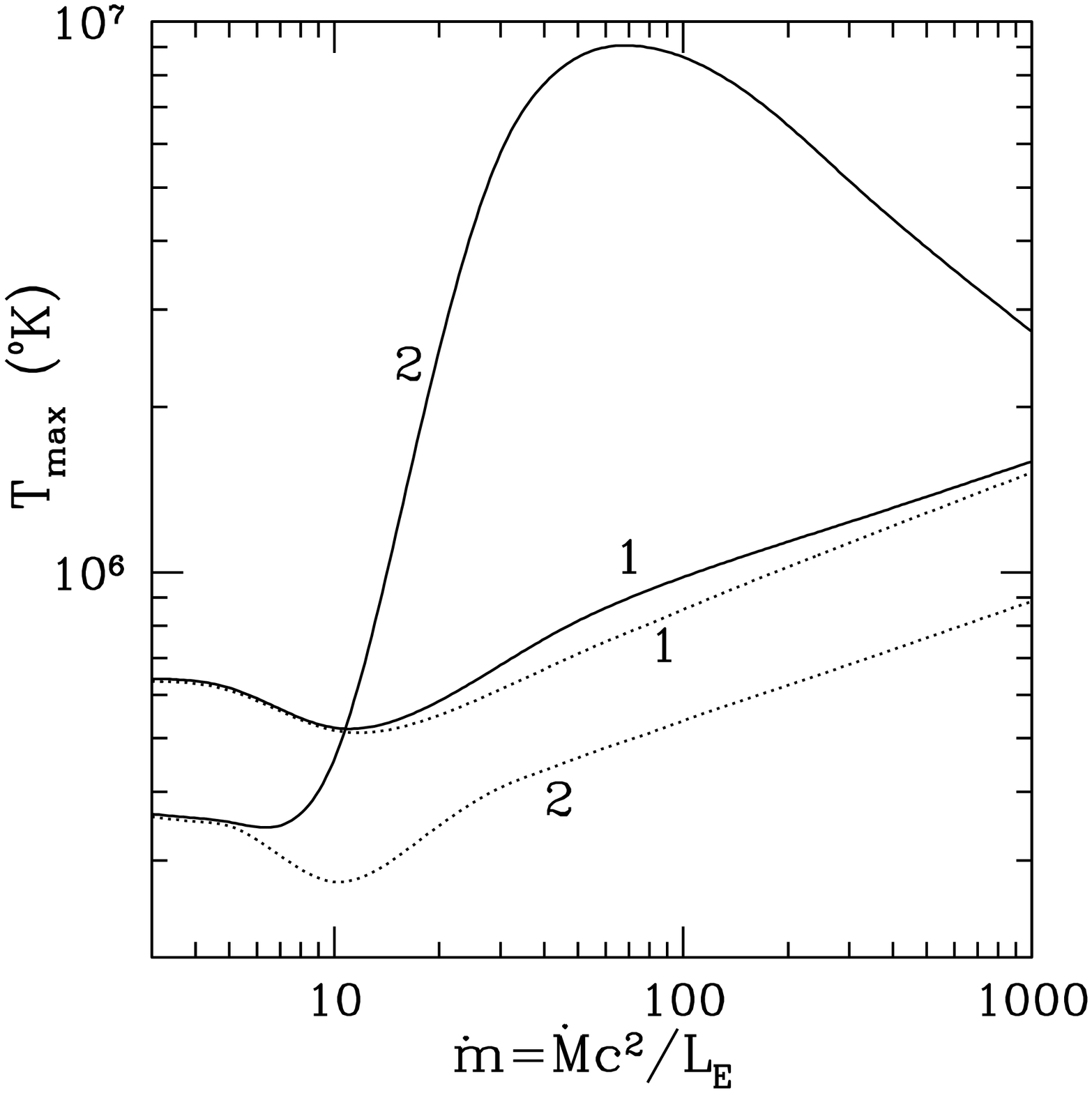,width=0.5\textwidth}
\psfig{figure=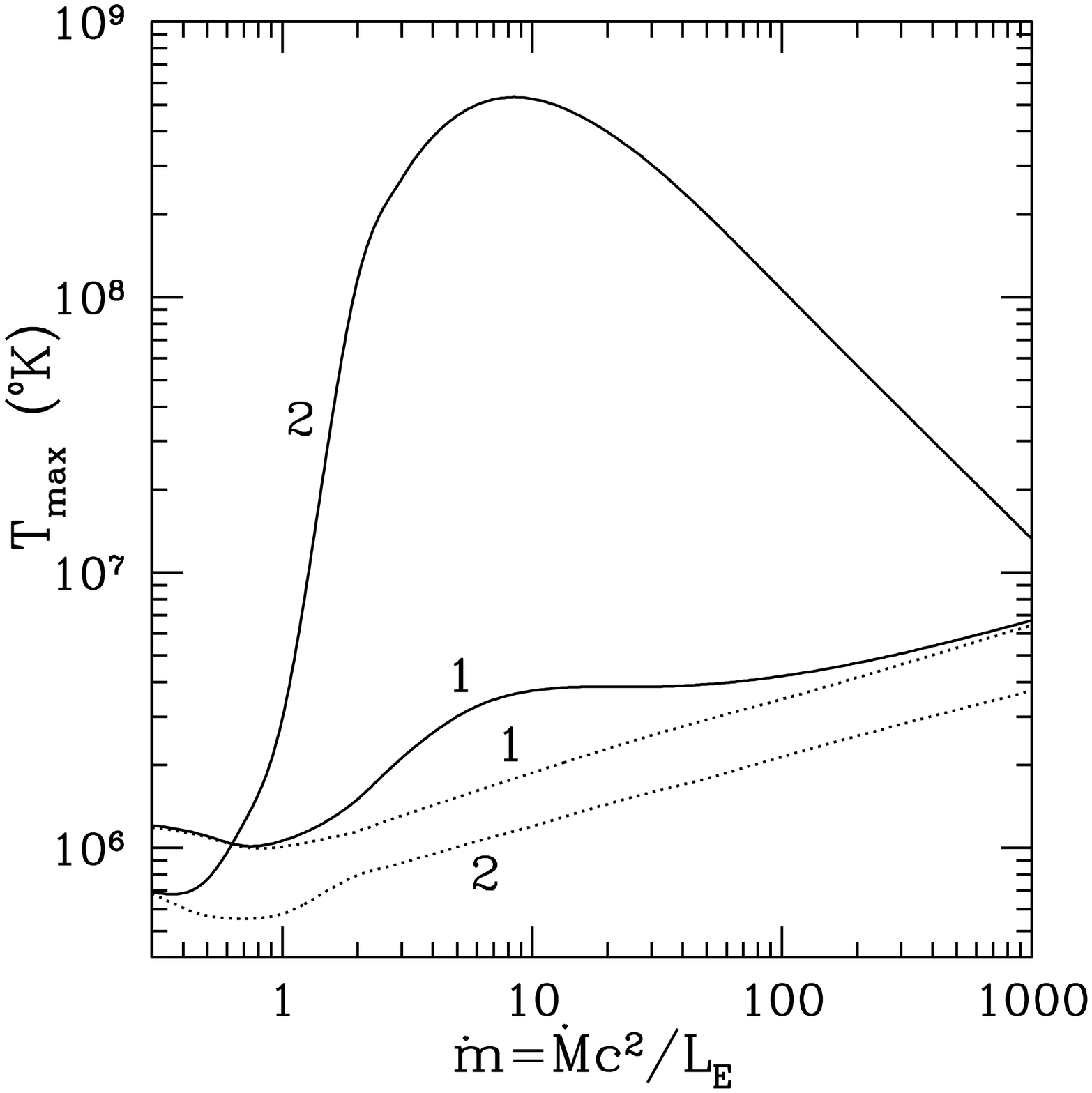,width=0.5\textwidth}
}
\caption{ The temperature in the innermost region of the disk versus
$\dot{m}$ for Schwarzschild ($a_*=0$, $\dm_{\rm cr}=17.5$, {\it left panel}) 
and Kerr ($a_*=0.998$, $\dm_{\rm cr}=3.11$, {\it right panel}) black holes 
in the cases of $\alpha=0.03$ (marked 1) and $\alpha=0.3$ (marked 2). 
A black hole mass $M=10^8M_\odot$ is assumed.
The dotted curves show the blackbody temperature $T_{\rm eff}$.
(From Beloborodov 1998a.)
} 
\label{fig-1}
\end{figure}
%%%%%%%%%%%%%%%%%%%%%%%%%%%%%%%%%%%%%%%%%%%%%%%%%%%%%%%%%%%%%%%%%%%%%%%%%%%
%\begin{center}
%\leavevmode
%%\epsfxsize=8.4cm \epsfbox{fig1.ps}
%\epsfxsize=9cm \epsfbox{fig1a.ps} \epsfxsize=9cm \epsfbox{fig1b.ps}
%\end{center}

The possibility of the overheating in near-critical disks was 
pointed out by Shakura \& Sunyaev (1973) and discussed later (e.g., 
Liang \& Wandel 1991; Bj\"ornsson et al. 1996). 
In the standard model, the disk density scales as $n\propto\dm^{-2}$, while the 
heating rate $F^+\propto\dm$. The resulting temperature of the overheated gas 
increases with $\dm$. The slim disk model allows one to follow this tendency 
toward 
the super-critical regime. Figure 1 shows the results for a massive black hole
in an AGN. The strongest overheating occurs at $\dm\sim 3-4$ $\dmcr$.
At $\dm\gg\dmcr$ the density increases, $n\propto\dm$, and the temperature 
falls down. A similar overheating occurs in GBHs ($M\sim 10 M_\odot$) if
$\alpha>0.1$.

The main process cooling the plasma in overheated disks with large $\dM$ 
is the saturated Comptonization of bremsstrahlung photons (cf. Rybicki \& 
Lightman 1979). The plasma temperature is determined by the heating=cooling 
balance (Beloborodov 1998a), 
\beq
\dot{w}^+\equiv\frac{F^+}{2H}\approx\dot{w}_{\rm ff}\left[\frac{w}{w_{\rm pl}}
           + A\left(1-\frac{w}{w_{\rm pl}}\right)\right]
            \left(1-\frac{w}{w_{\rm pl}}\right),
\eeq
where $\dot{w}_{\rm ff}=1.6\cdot 10^{-27}n^2\sqrt{T}$ is the free-free 
cooling rate, $w_{\rm pl}=a_rT^4$,
and $A$ is the Compton amplification factor.
In the optically thick limit ($w_{\rm pl}\approx w$) this equation yields
$\dot{w}^+=\dot{w}_{\rm ff}(1-w/w_{\rm pl})$, while in the overheated case
($w_{\rm pl}\gg w$) it transforms into $\dot{w}^+=A\dot{w}_{\rm ff}$.
The Compton amplification factor achieves $A\sim 300$ in the hottest models.
The efficient Compton cooling prevents a transition to an extremely
hot two-temperature state (see also Bj\"ornsson et al. 1996).   

The overheating will strongly affect the disk spectrum: the spectrum of a 
relativistic disk with $\dm\simgt\dmcr$ can extend to the hard X-ray band,
in contrast to the spectra of blackbody pseudo-Newtonian models 
(Szuszkiewicz, Malkan \& Abramowicz 1996).
%Hot disks with $\dm\simgt\dmcr$ may produce hard X-rays in luminous sources.

Many GBHs and AGNs, however, accrete at 
sub-Eddington rates and their hard X-ray emission implies a different mechanism.

%%%%%%%%%%%%%%%%%%%%%%%%%%%%%%%%%%%%%%%%%%%%%%%%%%%%%%%%%%%%%%%%%%%%%%%%%%%

\section{Two-temperature disks}
   
\subsection{The SLE model}

To explain the hard spectrum of the classical black hole candidate Cyg X-1, 
Shapiro, Lightman \& Eardley (1976, hereafter SLE) suggested a two-temperature
model of an accretion disk with protons much hotter than electrons, 
$T_p\gg T_e$. The hot proton component then dominates the pressure and keeps 
the disk geometrically thick. This in turn leads to 
% The sound speed, $c_s=kT_p/m_p$, 
% is a sizable fraction of the Keplerian velocity. The disk has 
a low density which scales $\propto H^{-3}$ for given $\dm$ and $\alpha$. 
A low density implies a low rate of Coulomb energy exchange between the 
protons and the electrons and allows one to keep $T_p\gg T_e$ self-consistently. 
%a result, Coulomb collisions are not able to cool the protons down to the 
%electron temperature. 

The four main equations of the disk structure are the same as in the standard 
model (see Section 2). The SLE disk has two temperatures. Therefore, 
the heating=cooling balance is now described by two equations: 

{\it i) Energy balance for the protons.} The model assumes that viscous 
dissipation heats preferentially the protons and then the 
bulk of this energy is transferred to the electrons through Coulomb collisions, 
i.e.,
$$
  \frac{F^+}{2H}=\frac{nkT_p}{t_{ep}}, \qquad {\rm where} \qquad
   t_{ep}\approx 17 \frac{T_e^{3/2}}{n}
$$
is the Coulomb cooling time (with a Coulomb logarithm $\approx 15$, see 
Spitzer 1962). This equation yields
\begin{eqnarray}
 \theta_p\theta_e\approx 10^{-2}\frac{\dm^{2/3}}{\alpha^{4/3}}
                         \left(\frac{r}{r_g}\right)^{-1}S^{-2/3},
\end{eqnarray}
where $\te\equiv kT_e/m_ec^2$ and $\tp\equiv kT_p/m_pc^2$.

{\it ii) Energy balance for the electrons.} The electrons cool
mainly by upscattering soft radiation coming from an outer cold disk or
from dense cloudlets embedded in the hot flow. 
The cooling proceeds in the regime of 
unsaturated Comptonization (cf. SLE). In this regime, 
the disk parameters adjust so that $y=4\te\max(\tau,\tau^2)\approx 1$.
Typically, $\te\simlt 1$ and $\tau\sim 1$. The condition $y\approx 1$ gives
\begin{eqnarray}
  \frac{\tp}{\te}\approx\frac{\sqrt{2}\dot{m}}{\alpha}
  \left(\frac{r}{r_g}\right)^{-3/2}
  \; \mbox{if} \; \tau\simlt 1 \qquad \mbox{and} \qquad 
  \frac{\tp^2}{\te}\approx\frac{\dm^2}{2\alpha^2}
  \left(\frac{r}{r_g}\right)^{-3} \; \mbox{if} \; \tau>1.
\end{eqnarray}

Equations (10-11) yield an electron temperature that weakly depends on 
radius, $\te\approx 0.1 (\alpha\dm)^{-1/6}(r/r_g)^{1/4}$.
The complete set of equations allows one to express the other 
disk parameters as functions of radius (see SLE). In particular, the proton 
temperature and the disk height are
$(H/r)^2\sim\tp/\theta_{\rm vir}\sim 0.1 \alpha^{-4/3}\dm^{2/3}(r/r_g)^{-1/4}$
where $\theta_{\rm vir}\sim r_g/r$ is the virial temperature. 

The model was further developed to include effects of pair production
(e.g., Liang 1979b; Bj\"ornsson \& Svensson 1992; Kusunose \& Mineshige 1992),
a non-thermal particle distribution (Kusunose \& Mineshige 1995), and 
cyclosynchrotron radiation (Kusunose \& Zdziarski 1994).
A relativistic version of the SLE disk around a Kerr black hole was calculated
by Bj\"ornsson (1995).

The SLE model is in agreement with observed spectra of GBHs and 
AGNs. The shortcoming of the model is that it is thermally unstable 
(Pringle 1976; Piran 1978): the assumed energy balance would be destroyed by 
a small perturbation of the proton temperature. An increase in 
$T_p$ would result in disk expansion in the vertical direction.
Then the Coulomb cooling is reduced (and the heating rate is increased) 
and $T_p$ increases further, leading to an instability.

\subsection{The ADAF model}

In the SLE model only a fraction of the dissipated energy remains in the proton 
component to keep it hot, and the bulk of the energy is passed to 
the electrons and radiated.
Ichimaru (1977) considered a disk of so low density that 
the heated protons are unable to pass their energy to the electrons on the 
time-scale of accretion. The protons then accumulate the energy and advect it. 
Such a flow is thermally stable. It emits only a fraction of the 
dissipated energy that has been passed directly to the electrons. 

If viscous dissipation heats mostly the protons, 
then the radiative efficiency of the accretion flow is very low, and
the energy is either advected into the black hole (Ichimaru 1977; 
Rees et al. 1982) or transported outward in the
form of a hot wind (Blandford \& Begelman 1998, see also Begelman, these 
proceedings).  The former case (ADAF) may be 
modeled in 1D approximation (see Narayan et al. 1998 for a review). 
The latter case requires detailed 2D simulations 
(Igumenshchev \& Abramowicz 1998) which are much more difficult:
the very formulation of a 2D problem implies
additional assumptions (local viscosity prescription, local heating rate,
and 2D boundary conditions).

The assumption that protons are heated preferentially has recently been 
assessed 
%Bisnovatyi-Kogan \& Lovelace 1997; Gruzinov 1998; Quataert 1998; 
(see Quataert \& Gruzinov 1998; Quataert, these proceedings, and references 
therein). 
It was argued that the dissipation of Alfv\'enic turbulence in the disk
may heat mostly
protons only if the magnetic energy density is $\simlt 10$\% of 
the proton pressure. If the magnetic field is the source of viscosity, 
one expects $\alpha <0.1$ in a two-temperature disk.

The necessary condition for ADAF models is that the time-scale for 
Coulomb cooling, $t_{ep}$, exceeds 
the accretion time-scale, $t_a\approx (\alpha\OmK)^{-1}$. It requires
$$
    \dot{m}\simlt 10^2 \alpha^2 \te^{3/2}.
$$
The typical electron temperature in the calculated ADAF models is $\sim 10^9$ K
(e.g., Nakamura et al. 1997; Esin et al. 1998), 
and the corresponding maximum accretion rate is 
\begin{equation}
     \dot{m}_{\rm max}\sim 10 \alpha^2. 
\end{equation}

From hydrodynamical point of view, the two-temperature ADAFs are 
similar to the super-critical disks. Again, most of the dissipated 
energy is stored inside the disk and it swells up to a quasi-spherical shape. 
The radial pressure gradients and deviations from Keplerian rotation are
dynamically important, and the full set of differential equations should be 
solved to obtain a solution for the disk structure
(Chen, Abramowicz \& Lasota 1997; Narayan, Kato \& Honma 1997). 
At large distances,
$r\gg r_g$, the ADAF parameters have a power-law dependence on radius
(Narayan \& Yi 1994). An approximate description of 
the advective disk was given by Abramowicz et al. (1995) who assumed that 
the disk rotates with a Keplerian velocity in the pseudo-Newtonian potential.
Recently, relativistic solutions have been calculated
for ADAFs in Kerr geometry (Abramowicz et al. 1996; 
Peitz \& Appl 1997; Igumenshchev, Abramowicz \& Novikov 1998; 
Popham \& Gammie 1998) and the expected spectra have been 
discussed (Jaroszy{\'n}ski \& Kurpiewski 1997).
The ADAF model is applied mainly to low 
luminosity objects with suspected black holes, such as Sgr A$^*$,
nuclei of elliptical galaxies, and X-ray novae in quiescent state 
(see Narayan et al. 1998 and references therein).

The possibility of a two-temperature accretion flow in Cyg X-1 and other GBHs
in the hard state has again been addressed by Narayan (1996), Esin et al. 
(1998), and Zdziarski (1998). Their basic picture of accretion is the same as 
in the SLE model: at some radius $\rtr$, the standard ``cold'' disk undergoes 
a transition to a hot two-temperature flow which emits hard X-rays by 
Comptonizing soft radiation in the unsaturated regime. In contrast to SLE, 
radial advection of heat is taken into account, and the parameters
($\dm$ and $\alpha$) are chosen so that advection and radiative cooling are
comparable. Then the flow is a good emitter and at the same time it may be 
stabilized by advection. 
It implies that the accretion rate is just at the upper limit (eq. [12]),
and $\alpha\sim 0.3$ has to be assumed, as typically $\dm\sim 1$ in GBHs. 
The optical depth of the flow near the black hole is approximately
$$
   \tau\sim \frac{\dot{m}}{\alpha}\sim 1.
$$ 
Combined with the typical electron temperature $T_e\sim 10^9$ K,
this yields $y\sim 1$, just what is needed to be in the regime of unsaturated 
Comptonization that is believed to generate the X-ray spectrum. 

An advantage of a hot disk model is its consistency with the observed weak 
X-ray reflection in GBHs in the hard state (Ebisawa et al. 1996; 
Gierli\'nski et al. 1997; Zdziarski et al. 1998; \.Zycki, Done \& Smith 1998). 
Observations suggest that cold gas reflecting the X-rays covers a modest solid 
angle $\Omega\sim 0.3\times 2\pi$ as viewed from the X-ray source. This is 
consistent with the outer cold + inner hot flow geometry (for alternative 
models see Section 6.4). Provided the two conditions 
$\dm\approx\dm_{\rm max}$ and $\tau\sim 1$ are satisfied, and 
the inner radius of the cold disk, $\rtr$, 
is properly chosen, the advective model may approximately describe
the hard state of GBHs. A small increase ($\sim 10$ \%) in $\dm$ would then
result in a transition to the soft state, as the hot flow collapses to 
a standard thin disk (Esin et al. 1998). A similar small decrease in $\dm$
would lead to transition to a low-efficiency ADAF which may be associated
with the quiescent state in transient sources.

There is then a question, why the hard state is so widespread if it 
requires the fine-tuning of both $\dm$ and $\alpha$? In particular, why
the hard state is stable in Cyg X-1 while its luminosity varies by a factor
of $\sim 2$? 
Variations in luminosity are presumably due to variations in $\dm$ and, 
according to the ADAF model, the change in $\dm$ must 
switch off the hard state and cause a transition to 
the soft or quiescent state. The latter has never been observed in Cyg X-1. 
Moreover, the observed slope of the X-ray spectrum in the hard state is kept 
constant with varying luminosity (Gierli\'nski et al. 1997), which indicates a 
persistent $y$-parameter of the emitting plasma.

One possible explanation is that the disk has $\dot{m}>\dm_{\rm max}$, 
being composed of two phases, hot and cold. 
The thermal instability of a hot disk at $\dm>\dm_{\rm max}$ may
continue to produce a cold phase until $\dm$ in the hot phase is
reduced to $\dm_{\rm max}$. Then a balance between the cold and hot
phases might keep both conditions $\dm\approx\dm_{\rm max}$ and 
$\tau\sim 1$ for the hot phase (Zdziarski 1998).
% The remaining accreting mass should then concentrate in cold clumps. 
Hot accretion disks with cold clumps of gas have recently
been discussed in different contexts (Kuncic, Celotti \& Rees 1997;
Celotti \& Rees, these proceedings; Krolik 1998; Krolik, these proceedings). 
The cold clumps may also play a role as a source of seed soft photons for 
Comptonization (e.g., Zdziarski et al. 1998).

The dynamics of a cold phase produced in the hot disk is, however, a very 
complicated and unresolved issue. In particular,  
it is  unclear whether the cold 
phase may accrete independently with its own radial velocity,
or if it rather is ``frozen'' in the hot phase. 
The cold clumps moving through the hot medium should be violently 
unstable, and a model for dynamical equilibrium is needed for a
continuously disrupted/renewed cold phase. If the clump life-time
exceeds the time-scale for momentum exchange with other clumps and/or with 
the hot medium, then the clumps should settle down to the equatorial plane 
and form a cold thin disk. One then arrives at the disk-corona model.

%%%%%%%%%%%%%%%%%%%%%%%%%%%%%%%%%%%%%%%%%%%%%%%%%%%%%%%%%%%%%%%%%%%%%%%%%%%%%%

\section{The disk-corona model}

\subsection{Magnetic flares}

That the observed X-rays can be produced in a hot corona of a 
relatively cold accretion disk extending all the way to the black hole 
was suggested a long time ago (e.g., Bisnovatyi-Kogan \& Blinnikov 1977; 
Liang 1979a). Most likely,
a low density corona is heated by reconnecting magnetic loops 
emerging from the disk (Galeev, Rosner \& Vaiana 1979, hereafter GRV). 
This implies that the corona is coupled to the disk by the magnetic field
(for alternative models, where the corona accretes fast above the disk, see, 
e.g., Esin et al. 1998; Witt, Czerny \& \.Zycki 1997; Czerny et al., these 
proceedings).

The usually exploited model for the corona formation is that of GRV.
%Galeev et al. (1979, hereafter GRV). 
According to the model,
a seed magnetic field is exponentially amplified in the disk due to
a combination of the differential Keplerian rotation and the turbulent
convective motions. The amplification time-scale at a radius $r$
is given by $t_{\rm G}\sim r/3v_c$ where $v_c$ is a convective 
velocity. GRV showed that inside luminous disks the field is not able to 
dissipate at the rate of amplification. Then buoyant magnetic loops 
elevate to the corona where the Alfv\'enic velocity is high and the magnetic 
field may dissipate quickly.

The coronal heating by the GRV mechanism is, however, not sufficient to
explain the hard state of GBHs (Beloborodov 1999).
The rate of magnetic energy production per unit area of the disk equals
$F_B=2H\overline{w}_B/t_{\rm G}$ where $\overline{w}_B=\overline{B^2}/8\pi$ is 
the average magnetic energy density in the disk at a radius $r$. One can 
compare $F_B$ with the total dissipation rate, $F^+=3t_{r\varphi}c_{\rm s}$,
to get
$$
\frac{F_B}{F^+}=\frac{H}{r}\frac{t_{r\varphi}^B}{t_{r\varphi}}. 
$$
Here $\trf^B=\overline{B_\varphi B_r}/4\pi$ and we took into account that 
$B_\varphi/B_r=c_{\rm s}/v_{\rm c}$ in the GRV model.
Hence, the GRV mechanism is able to dissipate only a small
fraction $\sim H/r\ll 1$ of the total energy released in the disk. 
%This would be in conflict with the spectra of GBHs in the hard state, where 
%a large fraction of the energy is emitted in hard X-rays.

Recent simulations of MHD turbulence indicate that the magneto-rotational 
instability efficiently generates magnetic energy in the disk 
(see Balbus \& Hawley 1998). The
instability operates on a Keplerian time-scale, which is $\sim H/r$ times
shorter than $t_{\rm G}$, and it produces magnetic energy at a rate $F_B\sim F^+$.
%i.e., fast enough to explain the main energy release as magnetic 
%dissipation. 
A low dissipation rate inside the disk would lead to buoyant transport of 
generated magnetic loops to the corona (GRV). 
%Combined with the GRV argument for magnetic buoyancy, this yields that a large
%fraction of $F^+$ may dissipate in the corona. Then 
The hard state of an accretion disk may be explained as being due to 
such a ``corona-dominated'' dissipation. By contrast, in the soft state, 
the bulk of the energy is released inside the optically thick disk
%which radiates as a blackbody.
and the coronal activity is suppressed. 

In the hard state, the corona is the place to which magnetic stress driving
accretion is transported and released.  Conservation of angular momentum 
reads (see eq.[4])
$$
 2H\trf=\frac{\dot{M}}{2\pi}\OmK S.
$$
For a standard radiation-pressure-dominated disk it gives
$\trf\approx m_pc\OmK/\sT$. If
a large fraction, $\zeta$, of $F^+$ is dissipated above the cold disk, then
the disk height is reduced by a factor $(1-\zeta)$ (Svensson \& Zdziarski
1994). It implies that $t_{r\varphi}$ is increased by a factor
$(1-\zeta)^{-1}$. The magnetic corona is expected to be inhomogeneous and
the main dissipation probably occurs in localized blobs where the
magnetic energy density $w_B$ is much larger than the average $\trf$.
The accumulated magnetic stress in such a blob may suddenly be released 
on a time-scale $t_0\sim 10 r_b/c$ (the ``discharge'' time-scale, 
see Haardt, Maraschi \& Ghisellini 1994) where $r_b$ is the blob size. 
This produces a compact flare of luminosity $L\sim r_b^3 w_B/t_0$.

In this picture of the hard state, the binding energy of spiraling gas is 
liberated in intense flares atop the accretion disk. 
There is no detailed model for the magnetic flare phenomenon despite the fact 
that magnetic flares have
been studied for many years in the context of solar activity. In particular,
it is unclear whether the flaring plasma should be thermal or non-thermal.
Observations therefore play a crucial role in developing a model.
Observed X-ray spectra of black hole sources suggest that the bulk of emission 
comes from a thermal plasma with a typical temperature $kT\sim 50-200$ keV and 
Thomson optical depth $\tT\sim 0.5-2$ (see Zdziarski et al. 1997; Poutanen 
1998). The narrow dispersion of the inferred $T$ and $\tT$ indicates the 
presence of a standard emission mechanism. 

One possible scenario has been developed assuming a large compactness parameter
of the flares $l=L\sT/r_bm_ec^3\sim 10-10^3$.
Then the flare gets dominated by $e^\pm$ pairs
created in $\gamma-\gamma$ interactions (e.g., Svensson 1986).
The pairs produced tend to keep $\tT\sim 0.5-2$ and $kT\sim 50-200$ keV,
in excellent agreement with observations. Yet, it is also possible that
the flares are dominated by a normal proton plasma with $\tT\sim 1$.
Improved data should help to distinguish between the models. In particular, 
detection of an annihilation feature in the X-ray spectra would help.  
%Exact values of $T$ and $\tT$ depend on $l$ 
%and on whether there is a non-thermal $e^\pm$ tail. 
% A strongly localized $e^\pm$ flare with $l\sim 10^2-10^3$ may
% have a low temperature, $kT<100$ keV, observed in some GBHs.
%In the presence of a non-thermal $e^\pm$ tail, the flare may have even
%$kT\simlt 50$ keV observed in the case of GX 339-4 (Zdziarski et al. 1998). 

\subsection{Compton cooling}

A flaring blob cools mainly by upscattering soft photons.
We will hereafter assume that soft radiation comes
from the underlying disk of a temperature $T_s$. 
The additional source of seed soft photons due to the cyclo-synchrotron emission 
in the blob is discussed, e.g., by Di Matteo, Celotti \& Fabian (1997) and 
Wardzi\'nski \& Zdziarski (these proceedings). When a soft photon of 
initial energy $\ep_s$ passes through the flare, it acquires {\it on 
average} an energy $A\ep_s$, where $A$ is the Compton amplification factor.
$A$ may also be expressed as $A=(L_{\rm diss}+L_s)/L_s$ where $L_{\rm diss}$ is
the power dissipated in the flare and $L_s$ is the intercepted soft luminosity.

The produced X-rays have a power-law spectrum whose slope $\Gamma$ 
depends on the relativistic $y$-parameter of the blob,
$y=4(\te+4\te^2)\tT(\tT+1)$. Leaving
aside the effects of the (unknown) geometry of the blob, 
one may evaluate $\Gamma$ by modeling radiative
transfer in the simplest one-zone approximation in terms of an escape probability,
as done, e.g., in the code of Coppi (1992). 
We have calculated the photon spectral index $\Gamma$ using Coppi's code
(see Figure 2). Within a few percent $\Gamma$ follows a power law
\begin{equation}
 \Gamma\approx\frac{9}{4}y^{-2/9}.
\end{equation}
This empirical relation is simpler than the approximation of 
Pozdnyakov, Sobol \& Sunyaev (1979), 
$\Gamma\approx 1+ [2/(\te+3)-\log\tT]/\log(12\te^2+25\te)$. 

%%%%%%%%%%%%%%%%%%%%%%%%%%%%%%%%%%%%%%%%%%%%%%%%%%%%%%%%%%%%%%%%%%%%%%%%%%
%Figure~\ref{fig-1}.
\begin{figure}
\centerline{
\psfig{figure=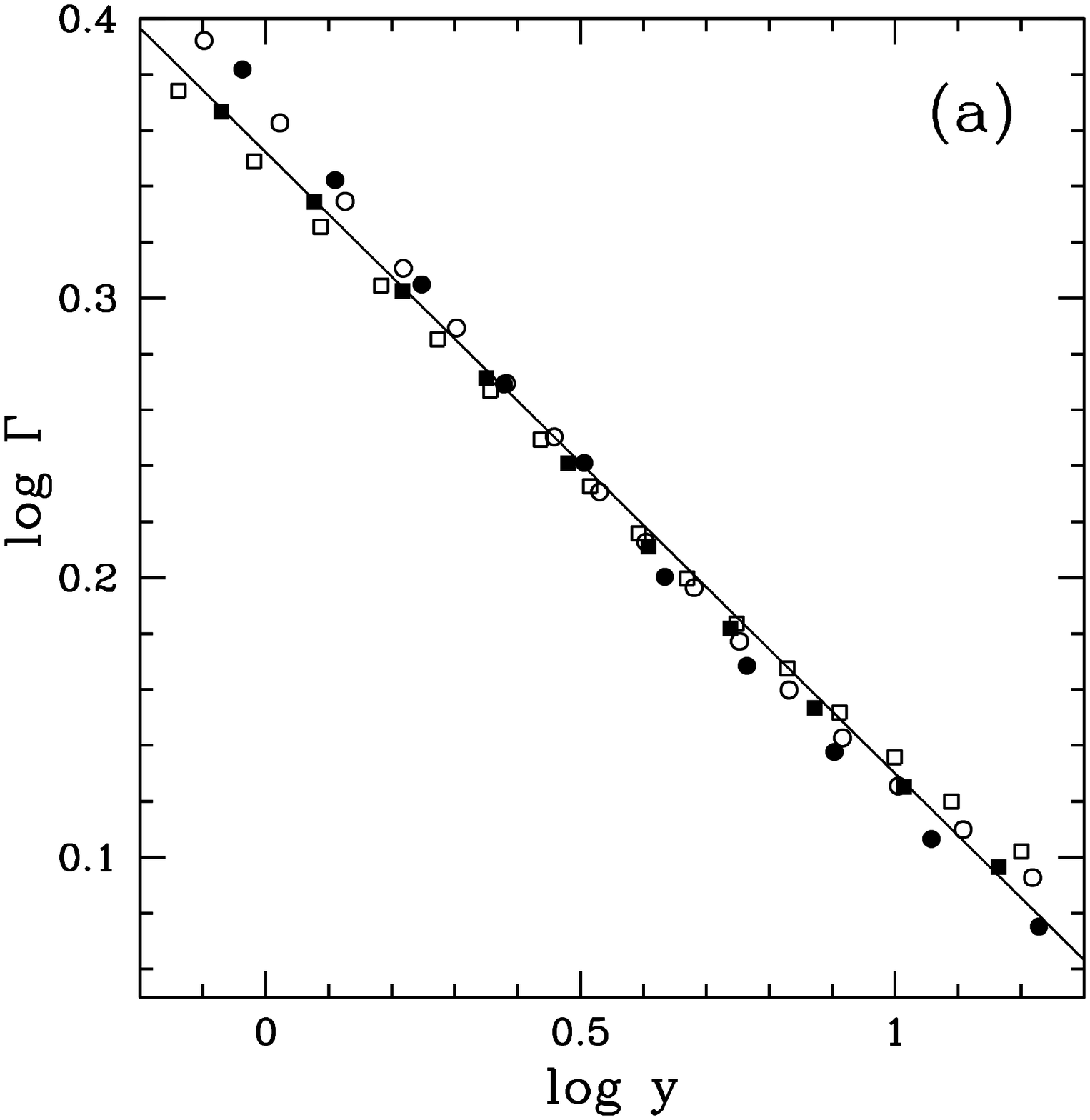,width=0.5\textwidth}
\psfig{figure=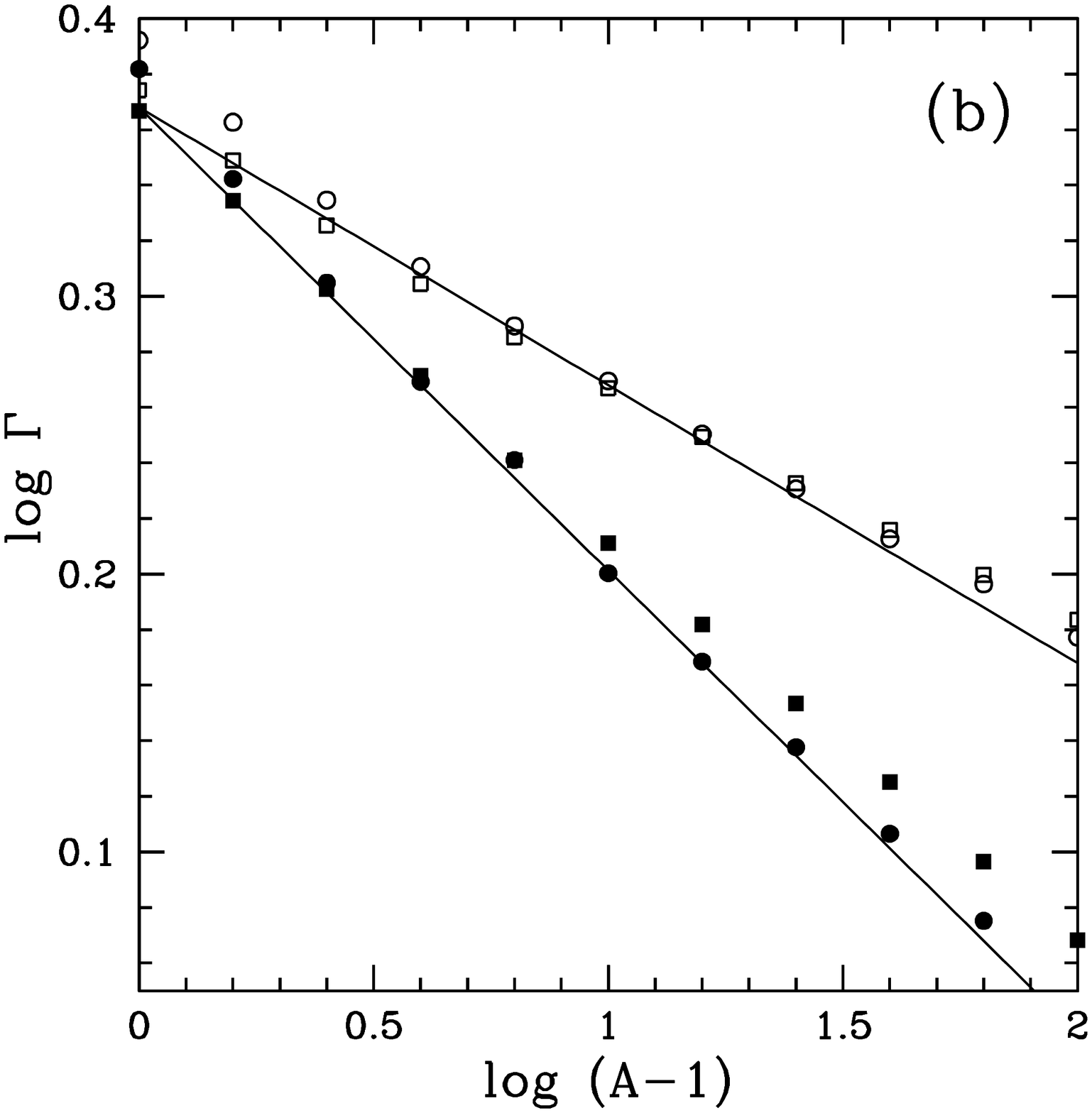,width=0.5\textwidth}
}
\caption{ (a) X-ray spectral index, $\Gamma$, versus $y$-parameter. 
%$y=4(\te+4\te^2)\tT(\tT+1)$. 
Open circles -- $kT_s=5$ eV, $kT=50$ keV,
open squares -- $kT_s=5$ eV, $kT=100$ keV, filled circles --
$kT_s=200$ eV, $kT=50$ keV, filled squares -- $kT_s=200$ eV, $kT=100$ keV.
The line shows the approximation (13).
(b) $\Gamma$ versus $A-1=L_{\rm diss}/L_s$. The two lines show 
the approximation (14) with $\delta=1/10$ and $\delta=1/6$.
} 
\label{fig-2}
\end{figure}
%%%%%%%%%%%%%%%%%%%%%%%%%%%%%%%%%%%%%%%%%%%%%%%%%%%%%%%%%%%%%%%%%%%%%%%%%%

One may also evaluate $\Gamma$ as a function of 
the Compton amplification factor. Then the result depends on $T_s$ 
which is typically a few $\times 10^6$ K in GBHs and a few $\times 10^4$ K 
in AGNs. The corresponding dependences $\Gamma(A)$ are shown in Figure 2b. 
To high accuracy ($\sim 3-4$ \%), the results can be approximated as 
\begin{equation}
\Gamma\approx\frac{7}{3} (A-1)^{-\delta},
\end{equation}
where $\delta\approx 1/6$ for GBHs and $\delta\approx 1/10$ for AGNs. 
Formula (14) is more accurate than the estimate of Pietrini \& Krolik 
(1995), $\Gamma\approx 1+1.6(A-1)^{-1/4}$, where the dependence 
on $T_s$ is neglected.

\subsection{The feedback of X-ray reprocessing by the disk}

The flares illuminate the underlying accretion disk that must 
reflect/reprocess the incident X-rays. The disk produces important features
in the observed X-ray spectrum, such as the Fe K$\alpha$ line and the Compton 
reflection 
bump, as extensively discussed in these proceedings. Even more important, 
the disk reprocesses a significant part of the X-ray luminosity into soft 
radiation. As the flare is expected to dominate the local disk emission,
the reprocessed radiation becomes the main source of soft photons.
Then the flares are ``self-regulating'': the flare temperature adjusts to keep 
$A=(L_s/L)^{-1}$ where $L_s/L$ is a fraction of the flare luminosity
that comes back as reprocessed radiation (Haardt \& Maraschi 1993;
Haardt et al. 1994). The resulting spectral slope is 
determined by the feedback of reprocessing, $L_s/L$. 
Detailed calculations of the predicted X-ray
spectrum were performed by Stern et al. (1995) and Poutanen \& Svensson (1996) 
and applied to Seyfert 1 AGNs (see Svensson 1996 for a review).

The calculated models are, however,
in conflict with observations of Cyg X-1 and similar black hole sources in 
the hard state (e.g., Gierli\'nski et al. 1997):
i) The observed hard spectrum corresponds to a Compton 
amplification factor $A\simgt 10$ and implies soft photon starvation 
of the hot plasma, $L_s\ll L$.
The model predicts $A\simlt 5$ unless the active blobs are elevated above 
the disk at heights larger than the blob size (Svensson 1996).
ii) The model with elevated blobs would yield a strong reflection component,
$R=\Omega/2\pi\approx 1$, where $\Omega$ is the solid 
angle covered by the cold matter as viewed from the X-ray source. 
The reported amount of reflection is small, $R\sim 0.3$. 

The weak reflection and soft photon starvation
may be explained if the cold reflector is disrupted near the black hole. 
This would agree with the idea that accretion proceeds as a hot two-temperature
flow in the inner region as discussed in Section 5. One may fit observed 
spectra with a toy model of a hot central cloud upscattering
soft photons supplied by a surrounding cold disk or by dense clumps inside
the hot region (e.g., Poutanen, Krolik \& Ryde 1997; Zdziarski et al. 1998).
The transient soft state is then explained as being due to shrinking of the 
hot region, so that the cold disk extends all the way to the black hole
and the bulk of the energy is dissipated inside the optically thick material
of the disk (Poutanen et al. 1997; Esin et al. 1998).

However, the weak reflection does not necessarily imply that the inner cold 
disk is disrupted in the hard state. One suggested alternative is that the
apparent weakness of the reflection features is due to a high ionization 
of the upper layers of the disk (Ross, Fabian \& Young 1998). 
 % A detailed analysis remains to be done to check whether the highly ionized 
 % reflector is consistent with observed spectra.
Another alternative is that the emitting hot plasma has a
bulk velocity directed away from the disk (Beloborodov 1999). 
Mildly relativistic bulk motion causes aberration reducing X-ray emission 
towards the disk. It in turn reduces the feedback of reprocessing 
and leads to a hard X-ray spectrum.

The coupling between the flare and the underlying disk can be approximately 
described assuming that the luminosity emitted downwards within an angle 
$\cos^{-1}\mu_s$ comes back to the flare. The effective $\mu_s$ depends on 
the flare geometry. 
E.g., a slab geometry of the active region corresponds to $\mus=0$, and 
an active hemisphere atop the disk has $\mus\approx 0.5$.
The feedback factor, $L_s/L$, of a flare of luminosity $L$ atop the disk 
is determined by three parameters:

\begin{itemize}

\item
The geometrical parameter $\mus$.

\item
The bulk velocity in the flare, $\beta=v/c$ (assumed to be perpendicular to 
the disk). 

\item 
The disk albedo, $a$. 
$\chi=1-a$ represents the efficiency of reprocessing of the incident X-rays.

\end{itemize}

In the static case ($\beta=0$), the reflection $R=1$ and $L_s=L\chi(1-\mus)/2$.
The corresponding amplification factor is $A=2/\chi(1-\mus)$.
With increasing $\beta$, $A$ increases, and $R$ is reduced.
The impact of a bulk velocity on $A$ and $R$ is summarized in Figure 3a for 
several $\mus$ and system inclinations $\theta$. In the calculations,
we assumed a typical albedo $a=0.15$ (e.g., Magdziarz \& Zdziarski 1995).
Note that the observed $R$ may be further reduced because the reflected 
radiation is partly upscattered by the blob.

One can evaluate the spectral index of a flare using equation (14).
For a typical $\mus\sim 0.5$ one gets 
$\Gamma\approx 1.9B^{-0.5}$ for GBHs and $\Gamma\approx 2B^{-0.3}$ for AGNs, 
where $B\equiv\gamma(1+\beta)$ is the aberration factor due to bulk
motion (Beloborodov 1999). E.g., for Cyg X-1
in the hard state, both the spectral slope $\Gamma\sim 1.6$ and the amount of 
reflection $R\sim 0.3$ can be explained assuming $\beta\sim 0.3$.

\subsection{The ejection model}

The inferred $\beta>0$ implies that the flares are accompanied by plasma 
ejection from the active regions, in contrast to the static corona model. 
In fact, one should expect bulk motion of a flaring plasma on theoretical side,
especially if the plasma is composed of light $e^\pm$ pairs. There is at least 
one reason for bulk acceleration: the flare luminosity, $L$, is partly 
reflected from the disk, and hence the flaring plasma is immersed in an 
anisotropic radiation field. The net radiation flux, $\sim L/r_b^2$, is 
directed away from the disk and it must accelerate the plasma. The transferred 
momentum per particle per light-crossing time, $r_b/c$, is $\sim l m_ec$ where
$l=L\sT/r_bm_ec^3$ is the compactness parameter of the flare. Hence, the
acceleration time-scale is $t_a\sim l^{-1}r_b/c$ for a pair plasma and
$t_a\sim (m_p/m_e)l^{-1}r_b/c$ for a normal proton plasma.
 % $f\sim F\sigma_{\rm T}/c$ is the accelerating radiative force.
The shortness of $t_a$ for a pair plasma implies that the pair bulk velocity
saturates at some equilibrium value limited by the radiation drag.
Using a simple toy model, one may estimate the expected velocity to be
in the range $\beta\sim 0.1-0.7$ (Beloborodov 1999).
A proton plasma may also be accelerated to relativistic velocities if
the flare duration exceeds $t_a$, which is quite probable.
 
%The net radiation flux, $F\sim L/r_b^2$, is directed away from the disk 
%and it must accelerate a pair plasma on a time-scale
%$t_a\sim m_ec/f\sim l^{-1}r_b/c$ where 
%$f\sim F\sigma_{\rm T}/c$ is the accelerating radiative force. 

%%%%%%%%%%%%%%%%%%%%%%%%%%%%%%%%%%%%%%%%%%%%%%%%%%%%%%%%%%%%%%
\begin{figure}
\centerline{
\epsfig{file=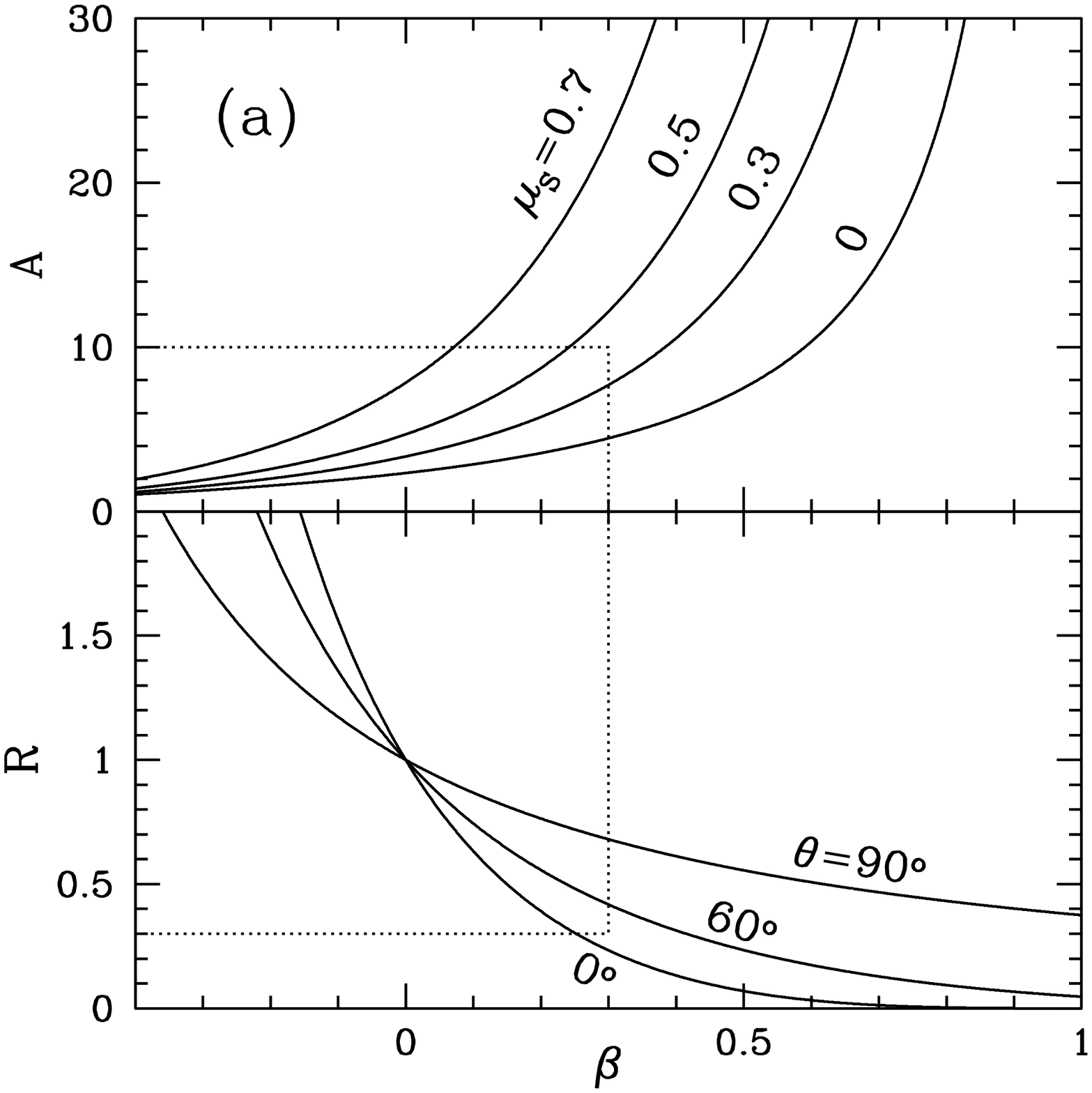,width=0.45\textwidth,height=0.52\textwidth}
\epsfig{file=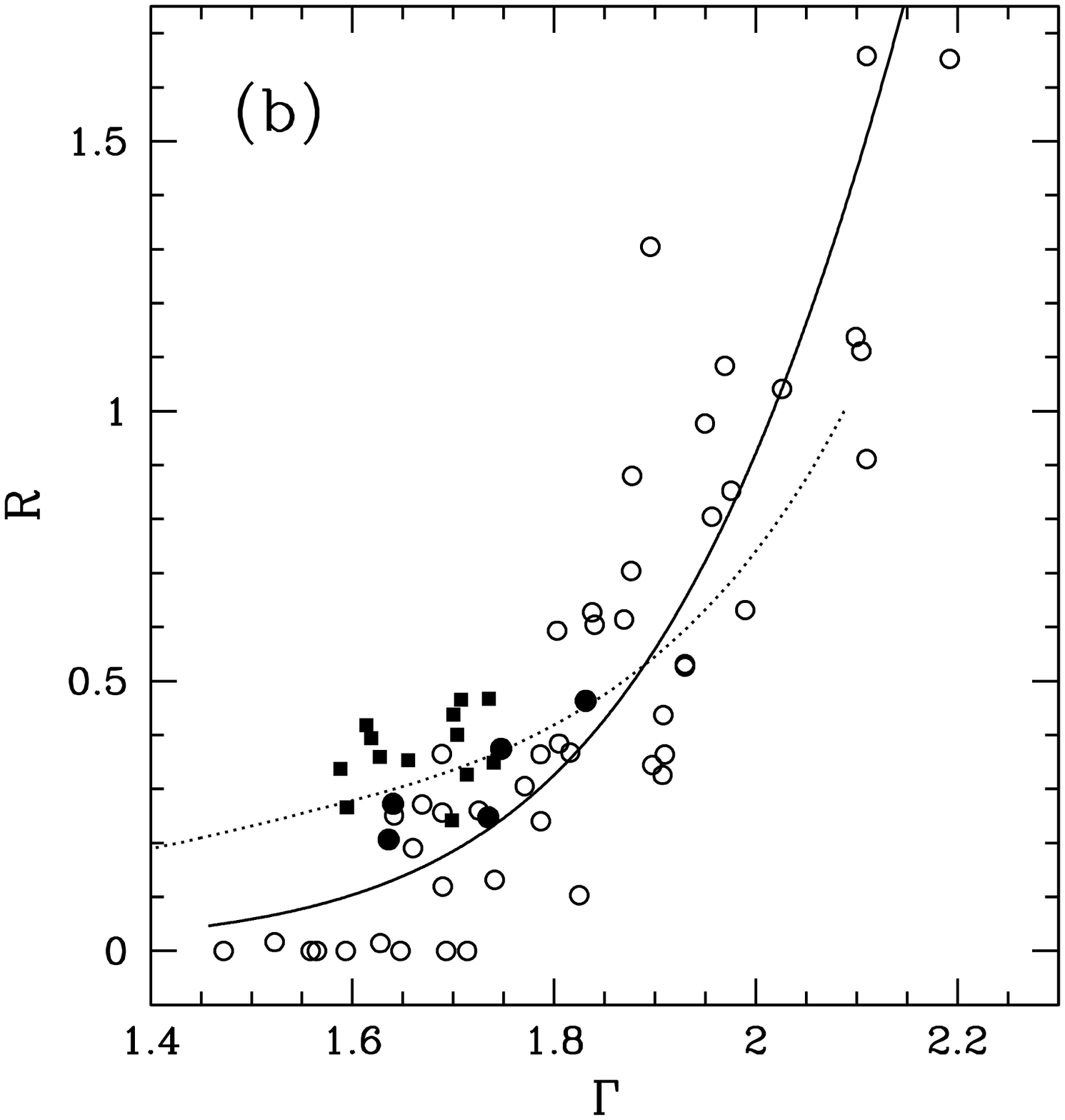,width=0.5\textwidth,height=0.52\textwidth}
}
\caption{ 
(a) {\it Top:} Compton amplification, $A$, of the reprocessed 
radiation as a function of 
 % the plasma bulk velocity, 
$\beta$.
A disk albedo $a=0.15$ is assumed. 
{\it Bottom:} Apparent amount of reflection, $R$.
The dotted line shows expected $A$ and $R$ for $\beta=0.3$. 
(From Beloborodov 1999.) $\;\;\;\;\;\;\;\;\;$
(b) Best fits to $\Gamma$ and $R$ for a number of AGNs (open circles), 
Cyg X-1 (filled squares), and GX 339-4 (filled circles).
The data are from Zdziarski et al. (1999). The solid curve shows the 
correlation $R(\Gamma)$ predicted by the ejection model for AGNs 
(assuming $\mus=0.55$ and inclination $\theta=45^{\rm o}$).
$R(\Gamma)$ predicted for AGNs by the hot disk model (see Zdziarski et al. 1999)
is shown by the dotted line.
} 
\label{fig-3}
\end{figure}
%%%%%%%%%%%%%%%%%%%%%%%%%%%%%%%%%%%%%%%%%%%%%%%%%%%%%%%%%%%%%%

The very magnetic dissipation may be accompanied by pumping a net momentum into
the flare at a rate $\sim L/c$. When the stored magnetic energy gets
released, the heated plasma may be ejected both toward and away from the disk.
Again, the large compactness parameter implies efficient momentum transfer to 
the plasma, $\sim lm_ec$ per particle per light crossing time. 
Possible ejection toward the disk corresponds to $\beta<0$.
 
Plasma ejection from magnetic flares is likely to occur in both GBHs and AGNs.
The plasma velocity may vary. An increase in $\beta$ leads to decreasing $R$ 
and $\Gamma$. A correlation between $R$ and $\Gamma$ is observed in GBHs and 
AGNs (Zdziarski, Lubi\'nski \& Smith 1999; Zdziarski, these proceedings) and 
it is well reproduced by the ejection model, see Figure 3b. The theoretical 
curve is plotted for AGNs, assuming a disk albedo $a=0.15$, 
$\mu_s=0.55$, and inclination $\theta=45^{\rm o}$.  These parameters are 
probably the most representative. The dispersion of data around the curve 
might be due to a dispersion in $\mus$ and inclination angles. The velocity, 
$\beta$, varies from $-0.2$ to $0.75$; $R=1$ corresponds to $\beta=0$.

The two model curves shown in Figure 3b are calculated for AGNs ($\delta=1/10$ 
in eq. [14]). For GBHs, $\Gamma$ should be systematically smaller as a result 
of a higher energy of seed soft photons. This tendency is seen in Figure 3b:
the Cyg X-1 and GX 339-4 data are shifted to the left as compared to the AGN 
data.

Outflows are usually expected in radio-loud objects where they have been
considered as a possible reason for weak Compton reflection (Wo\'zniak et al. 
1998). Figure 3b may indicate that plasma acceleration in X-ray coronas
of accretion disks is a common phenomenon in black hole sources. Note also 
that fast outflows may manifest themselves in optical polarimetric observations
of AGNs (Beloborodov 1998b; Beloborodov \& Poutanen, these proceedings).

%%%%%%%%%%%%%%%%%%%%%%%%%%%%%%%%%%%%%%%%%%%%%%%%%%%%%%%%%%%%%%%%%%%%%%%%%%%%%%

\section{Concluding remarks}

Disk-like accretion may proceed in various regimes.
Any specific model is based on assumptions including prescriptions for the
effective viscosity, the vertical distribution of the ion/electron heating, 
the energy transport out of the disk, etc.  
Large uncertainties in these prescriptions allow one to produce a lot of 
models, and observations should help to choose between them. 
Hard X-ray observations play an important role in this respect. They indicate
that a large fraction of the accretion energy is released in a rarefied plasma 
(corona) where the X-rays are generated by unsaturated Comptonization of soft 
photons. In most cases, the accreting mass is expected to be concentrated in a 
``cold'' phase, probably forming a thin disk embedded in the corona. 
Its gravitational energy transforms into the magnetic energy that is 
subsequently released in the corona.

Most likely, the dissipation mechanism is related to highly non-linear MHD 
which usually produces inhomogeneous and variable dynamical systems. 
In particular, the energy release in the corona is likely to proceed in 
magnetic flares generating observed temporal and spectral X-ray variability.
Idealized hydrodynamical models are unable to describe this process.
The difficulty of the problem is illustrated by the activity of the Sun, where 
theoretical progress is modest despite the fact that detailed observations are 
available. In accretion disks, magnetic fields are likely to play a crucial 
dynamical role and various plasma instabilities may take place. 
The instabilities probably govern the distribution of the plasma density and 
the heating rate. 

Given the difficulty of the accretion physics, studies of phenomena having
specific, observationally testable implications, are especially useful.
In particular, the reflection and reprocessing of the corona emission by
the cold disk provides diagnostics for accretion models. We here discussed a
plausible phenomenon -- bulk acceleration of the flaring plasma in the corona 
-- that strongly affects the reflection pattern and the radiative
coupling between the corona and the cold disk.

\acknowledgments

I thank J. Poutanen, R. Svensson, and A. A. Zdziarski for discussions
and comments on the manuscript. I am grateful to Andrzej Zdziarski for providing
the $R-\Gamma$ correlation data prior to publication and to Paolo Coppi for
his {\sc EQPAIR} code. 
This work was supported by the Swedish Natural Science Research Council
and RFFI grant 97-02-16975.

\end{document}